March 13, 2020

# Possibility of Disinfection of SARS-CoV-2 (COVID-19) in Human Respiratory Tract by Controlled Ethanol Vapor Inhalation

Tsumoru Shintake

Professor in Physics at OIST Graduate University
Okinawa Institute of Science and Technology Graduate University
1919-1 Tancha, Onna-son, Kunigami-gun, Okinawa, Japan 904-0495

Viruses such as SARS-CoV-2 and Influenza are lipophilic, enveloped viruses, and are relatively easy to inactivate by exposure to alcohols. The envelope mainly consists of the lipid bilayer, taken from the host cells at assembly/budding stage of the viral life cycle. Therefore the constitution of the lipid bilayer should be common in all SARS, MERS and influenza viruses, even after mutations, and thus these closely-related viruses will be disinfected by exposure to ethanol with the same concentration. Existing experimental data indicate that an ethanol concentration of 30~40 v/v% is sufficient to inactivate Influenza-A viruses in solution[1,2,3].

The author suggests that it may be possible to use alcoholic beverages of 16~20 v/v% concentration for this disinfection process, such as Whisky (1:1 hot water dilution) or Japanese Sake, because they are readily available and safe (non-toxic). By inhaling the alcohol vapor at 50~60°C (122~140°F) through the nose for one or two minutes, it will condense on surfaces inside the respiratory tract; mainly in the nasal cavity. The alcohol concentration will be intensified to ~36 v/v% by this process, which is enough to disinfect the corona virus on the mucous membrane. In this situation, our respiratory tract essentially works as an alcohol distillation apparatus (a condenser). This method also provides more moisture into respiratory tract, and helps to clean the inside of the nasal cavity by stimulating blowing of the nose, and also makes the mucous escalator work actively so that the self-clearing mechanism in the trachea will remove viruses faster.

An alternative prompt method is also discussed. We use 40 v/v% whisky or similar alcohol, dripping on a gauze, inhale the vapor slowly at room temperature. This method works well for the front part of the nasal cavity. This is suitable for clinical workers, because they may need to use prompt preventative measures at any time.

1. Introductions

Alcohol for disinfection (rubbing alcohol in US, Surgical spirit B.P.) has been well established to sterilize various bacterias and disinfect viruses. Alcohols target the bacterial cell envelope, with resultant lysis of the cell and release of the cellular content. Also, it is well established that lipophilic, enveloped viruses are easier to deactivate by alcohols than the non-enveloped viruses[1]. Nobuji Noda[2] reported Influenza-A (USSR/92/97) was disinfected by a 1 minute exposure to ethanol with a concentration of 30~40 v/v%. Recently Akemi Shinmyo[3] reported careful experimental results on five different strains of Influenza-A, where all viruses were disinfected by 1 minute exposure to ethanol at 27-36 v/v% (by reading result with reduction ratio 1/100). We quickly performed the same test with a common laboratory E. coli strain (DH5alpha), and observed that the bacterium becomes non-viable at an ethanol concentration of 30 v/v%. E. coli is a gram negative bacteria, which also has a lipid outer layer and is thus sensitive to ethanol.

The novel coronavirus SARS-CoV-2 has recently emerged from China with a total of 113,702 laboratory-confirmed cases (as of March 9, 2020) [4]. COVID-19 has become a global health concern, and is causing huge economical impact due to freezing industrial activity closely linked with supply chains in Wuhan, China, and blocking traffics in many countries. There is no vaccine

shintakelab@oist.jp

yet developed, and standard recommendations to prevent infection spread include regular hand washing, and covering mouth and nose when coughing and sneezing. One of the problems is that some spread might be possible before people show symptoms. Another problem is the suspected longer incubation time associated with this virus. We urgently need an additional preventative measure, which can stop, or at least slow down, the spread of this novel coronavirus.

In this paper, the author discuss possibility of disinfection of SARS-CoV-2 in our respiratory tract via ethanol vapor inhalation.

2. Physics of Diffusion Phenomena

To use ethanol solution for disinfection, we have to understand the physics of diffusion and evaporation. When we spray the ethanol solution on a surface for disinfection, for example, a door knob, we have to be careful about the time-dependent decay of the ethanol concentration inside the water droplet. As seen in the later section, the vapor pressure of ethanol is much higher than that of water, thus ethanol evaporates faster, and the remaining solution becomes simple water quickly. The mean-free path of molecules in solution is roughly 0.1 nm at room temperature, and it will take more than $10^4$ times collisions for a molecule to travel a few micro-meters to reach the surface, even in a straight pass, thus the diffusion phenomena dominates. The speed of evaporation of ethanol can be estimated by the diffusion speed of ethanol molecules within a droplet. The simple 1D solution of the Fick's diffusion law[5] is

$$x^2 = 4Dt \quad (1)$$

where $D$ is the diffusion constant, which follows Einstein–Smoluchowski relation (kinetic theory)

$$D = \mu k_B T \quad (2)$$

where $\mu$ is the mobility, $k_B$ is the Boltzmann's constant, $T$ is the absolute temperature. For more detail, we need to discuss the evaporation speed, which is equal to the probability of a molecule escaping from the surface through enthalpy change in the Maxwell-Boltzmann distribution. This is outside of the scope of this paper - we simply assume the diffusion dominates. The diffusion constants of ethanol are listed in Table-1, and typical diffusion parameters are listed in Table-2.

|  | Diffusion Constant (25 °C) |
|---|---|
| Ethanol in water solution | $\sim 1 \times 10^{-9} m^2/sec$ |
| Ethanol vapor in air | $1.1 \times 10^{-5} m^2/sec$ |

Table-1 Diffusion constant of the ethanol molecule in water and air.

| Diffusion Time | Diffusion Distance Ethanol in Water Solution | Diffusion Distance Ethanol in Air |
|---|---|---|
| 0.01 sec | 6 $\mu m$ | 0.6 mm |
| 1 sec | 60 $\mu m$ | 6 mm |
| 1 minute | 490 $\mu m$ | 50 mm |

Table-2 Diffusion time and distance of the ethanol molecule in water and air.

An idea to use ethanol-water spray having 10 $\mu m$ droplet size, for inhalation purposes, does not work. The reason is that, as seen in Table-2, to travel 6 $\mu m$, it takes only 0.01 sec, and assuming 3 m/sec spray speed, the droplet after flying 30 mm in air, all ethanol will diffuse out. The droplets become water, containing only low density ethanol, and thus will not be effective for disinfection.

Thus, if we use the spray for ethanol disinfection experiments, the outcomes would fluctuate greatly, and always higher concentration would be required. Also, in practice, the ethanol spray should be avoided, because of possible fire accident.

Therefore, the author proposes inhalation of the ethanol vapor through the nose, which then condenses inside our respiratory tract, and thus disinfects the corona virus.

3. How to Sterilize in our Respiratory Tract

As shown in Fig. 1, we inhale the ethanol vapor of alcohol solution at 50~60°C through the nostrils using a tall cup (heat insulating styrofoam, or a wine glass). We fill ~30 ml Whisky, diluting with 30 ml hot water (90°C). Half amount of them still works. The ethanol concentration becomes ~20%, and temperature goes 50~60°C. The cup should be brought close to the nose (press the rim on to the lip), and the ethanol vapor should be inhaled. Mixing air and alcohol temperature change do not affect on the ethanol concentration after condensation inside nasal cavity, but it is better to keep air mixing as small as possible to minimize vapor condensation within the cup.

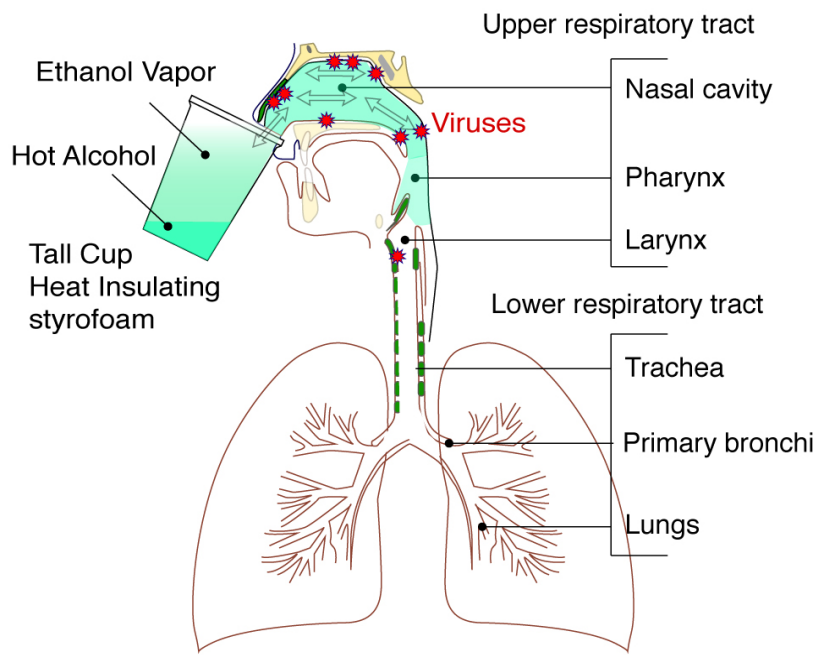

Fig. 1  Inhaling the alcohol vapor through the nostrils for disinfection of viruses inside the nasal cavity. Your respiratory tract as the alcohol distillation apparatus. This illustration is reproduced from original image found in Wikipedia: https://en.wikipedia.org/wiki/Trachea

It will be better restrict our breathing to shallower breaths, in order to reduce alcohol condensation in the trachea, where the important cilia-cells cover the surface and work for the mucous escalator as the self-clearing mechanism of the airways. The mucous thickness in trachea is only 10 $\mu m$[6], and it takes only 0.01 sec for ethanol to reach the epithelium layer as shown in Table-2. In order to minimize the risk of inflammation due to ethanol vapor (while the ethanol concentration is lowered after traveling from nasal cavity to trachea), we have to limit the depth of breath.

Figure 2 shows an example of controlled inhalation cycle. This is determined by the following factors.

(1) Normally the breath volume of adult is 500 ml/ breath × 12 breaths/min [7].

(2) The so-called "dead space", the volume of air that does not take part in the gas exchange, is typically 150 ml. If the breath is smaller than dead space, ethanol does not go into the lung and is not transferred into the blood.

(3) The volume of the nasal cavity of an adult is typically 30~37 ml [8]. The breath volume must be larger than this value. We assume 100 ml volume in each cycle.

(4) The surface area of the nasal cavity is roughly 100 cm² [9], by taking account of nasal cavity volume, average gap should be 6~8 mm. The diffusion time of an ethanol molecule for half of this gap is roughly 0.3 sec in air estimated from Eq. (1). The inhalation cycle time has to be longer than 0.3 seconds, so that the ethanol molecules can reach the nasal surfaces. After this period, we have to refresh the ethanol vapor, because most of ethanol molecules should be adsorbed into the mucous membrane. The air flow inside the nasal cavity is turbulent, due to the narrow and complicated pathway, which also accelerates the molecule adsorption speed. For more detail, we have to run numerical simulations on air flow (fluid dynamics) and condensation (thermodynamics).

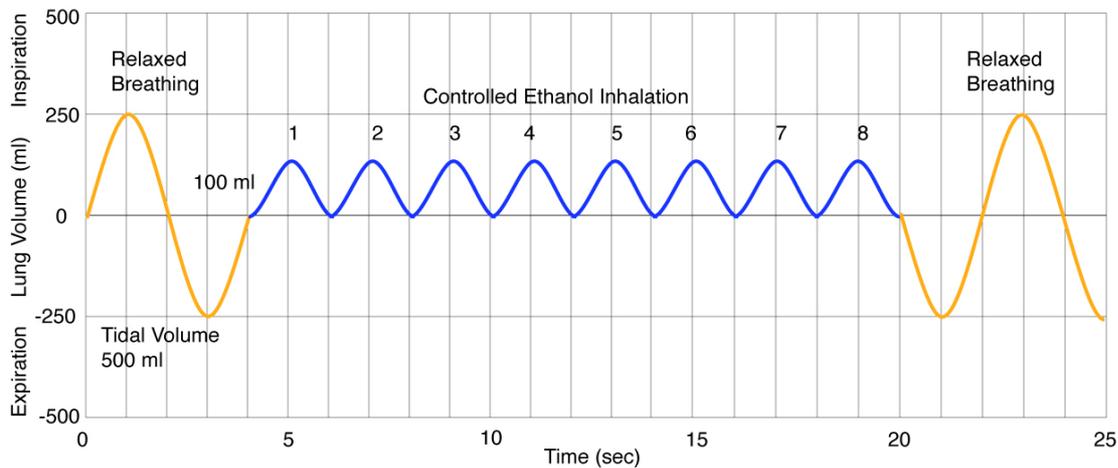

Fig. 2. Example of the controlled inhalation cycle. The vertical axis is the volume change of lung from the average volume at the relaxed breathing. During the inhalation, each breath should be around 100 ml, and eight repeats of shallow inhalation/exhalation through nose should be performed, followed by a few seconds of relaxed breathing outside the cup (providing oxygen). You may drink a small amount of the alcohol, cleaning around the pharynx, if permitted according to your age and local laws. We repeat this process for a few times, roughly one or two minutes in total. If you feel drunken, you have to control the breath shallower.

4. Ethanol Distillation Using Human Respiratory Tract

In this section, the pressures are written in $\times 10^5 Pa \approx 1\ atm$ (the standard atmosphere) unit. The alcohol evaporates faster than the water. In physics term, this is expressed in terms of vapor pressure. The vapor pressure of ethanol is higher than that of the water, that is why the alcohol distillation process works. We utilize this principle to bring the ethanol into our respiratory tract at higher concentration, efficiently and safely.

Before the detailed discussion, we have to note that the major gas component in the inhaling vapor of alcohol is still normal air (nitrogen and oxygen). These species can also exist in liquid form, but their boiling temperatures are very low ($N_2 = -195°C$, $O_2 = -183°C$), and they will thus not condense inside our body. Therefore we may neglect contributions from the air in the following discussions. The presence of the air makes mass transfer speeds slower, but the ethanol concentration in the condensed alcohol stays unchanged. We also neglect $CO_2$ in the expired gas, because it is a fairly small proportion. We treat only two components: ethanol and water in the following discussions. Table 3 summarizes the major physical properties of ethanol and water.

The vapor pressure of pure material is given by the following Antoine equation, which is a semi-empirical correlations describing the relation between vapor pressure and temperature for pure substances[5].

$$\log_{10} p^* = A - \frac{B}{C+T} \qquad (3)$$

where $p^*$ is the vapor pressure, $T$ is temperature in °C and $A$, $B$ and $C$ are component-specific constants. Those parameters are summarized in Table 4, and the vapor pressure curves for pure ethanol and water are shown in Fig. 3. At 78.4 °C, the vapor pressure of the ethanol reaches to the atmospheric pressure (1 atm), and thus boils.

|  | Ethanol | Water |
|---|---|---|
| Chemical Formula | $C_2H_5OH$ | $H_2O$ |
| Molar Mass | 46.1 g/mol | 18.0 g/mol |
| Density (at 40 °C) | 0.768 g/cm³ | 0.989 g/cm³ |
| Boiling Point | 78.24 °C | 99.97 °C |
| Heat of Vaporization (at 40 °C) (Enthalpy of vaporization) | 38.6 kJ/mol | 43.3 kJ/mol |

Table 3. Physical property of ethanol and water.

|  | A | B | C | $T_{min}(°C)$ | $T_{max}(°C)$ | 36.5(°C) |
|---|---|---|---|---|---|---|
| Ethanol | 10.329 | 1642.9 | 230.3 | -57 | 80 | $0.148 \times 10^5 P_a$ |
| Water | 10.196 | 1730.6 | 233.4 | 1 | 100 | $0.061 \times 10^5 P_a$ |

Table 4. Parameter for Antoine equation: T in °C and pressure in Pa.

The total pressure is given by Dalton's law,

$$P = \sum_{i=0}^{n} p_i \qquad (4)$$

where $p_i$ is the partial pressure of the component $i$ in the gaseous mixture, which relies on the ideal gas law;

$$p_i V = n_i RT \qquad (5)$$

where $V$ and $T$ are the volume and absolute temperature in Kelvin (°K), $R$ is the ideal gas constant. $n_i$ is the amount of substance. In the current discussions, the temperature change is roughly 20 °K, which is much smaller than room temperature (300 °K), thus we may assume temperature is constant. Because all gas components share the same volume, we can normalize eq. (5) as $V = RT$, finally we have

$$p_i = n_i \qquad (6)$$

The partial pressure is equal to the normalized amount of substance.
 The vapor pressure of mixture of liquids is given by Raoult's law. It states that the partial pressure of each component of an ideal mixture of liquids is equal to the vapor pressure of the pure component multiplied by its mole fraction in the mixture [5].

$$p_i = p_i^* x_i \qquad (7)$$

where $p_i$ is the partial pressure of the component $i$ in the gaseous mixture (above the solution), $p_i^*$ is the equilibrium vapor pressure of the pure component given by eq. (3), $x_i$ is the mole fraction of the component $i$ in the mixture (in the solution). The ethanol and water mixture is not an ideal solution, thus the actual vapor pressure shows negative or positive deviation due to the force acting between molecules. However, as we see later, the actual mole fraction in our alcohol is only 0.07, thus the deviation should be small, and we may treat our alcohol as an ideal solution.

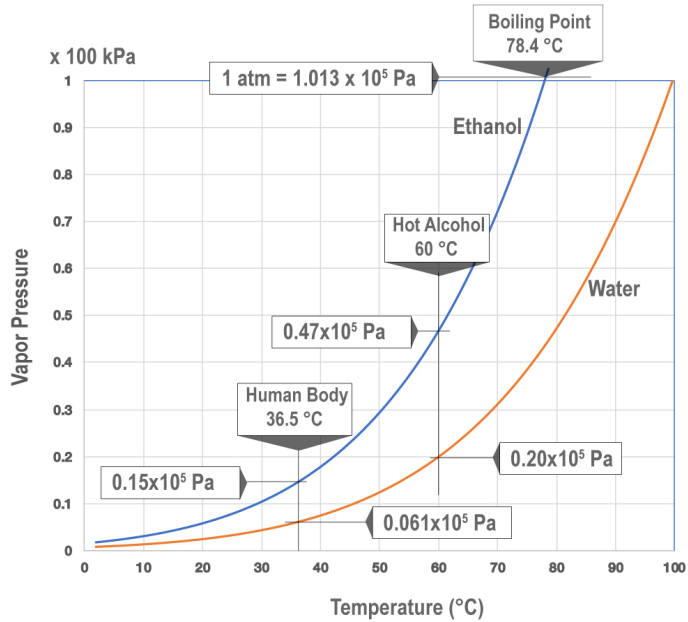

Fig. 3. Vapor pressure curve of the pure ethanol and water.

   Normally the concentration of alcohol is expressed using ABV: Alcohol By Volume, such as, 40 v/v%. Raoult's law says the partial pressure is proportional to the mole fraction. We need to convert ABV to mole-faction as follows.

$$x_1 = \frac{n_1}{n_1 + n_2} = \frac{m_1/46.1}{m_1/46.1 + m_2/18.0} = \frac{v_1 \rho_1/46.1}{v_1 \rho_1/46.1 + v_2 \rho_2/18.0} \qquad (8)$$

where $n_1, n_2$ are the mole contents of ethanol and water in unit volume. Using densities of $\rho_1 = 0.768$ and $\rho_2 = 0.989$, we have

$$x_1 = \frac{0.0167 v_1}{0.0167 v_1 + 0.0549 v_2} = \frac{0.304 v_1/v_2}{0.304 v_1/v_2 + 1.0} \qquad (9)$$

Using the definition of common ABV: Alcohol by volume (v/v%),

$$ABV = \frac{v_1}{v_1 + v_2}, \quad v_1/v_2 = \frac{ABV}{1 - ABV}. \qquad (10)$$

Finally, the mole fraction of ethanol is given by
$$x_1 = \frac{0.304\ ABV}{1 - 0.696\ ABV}, \quad ABV = \frac{x_1}{0.304 + 0.696\ x_1} \quad (11)$$

We use 20 v/v% alcohol (Whisky 1:1 water dilution), which is equal to $x_1 = 0.071$ mole fraction. The water mole fraction is $x_2 = 1 - x_1 = 0.93$, which means most of the molecules in solution are water. From Raoult's law, we find the partial pressures as follows. The mole contents are defined as shown in Fig. 4.

At high temperature $T_H = 60°C$;
  Ethanol: $n_{1H} = p_{1H} = p_{1H}^* x_1 = 0.47 \times 0.071 = 0.033$
  Water : $n_{2H} = p_{2H} = p_{2H}^* x_2 = 0.20 \times 0.93 = 0.19$

After inhaling into our body, the gas will be cooled down.

At low temperature $T_L = 36.5°C$;
  Ethanol: $n_{1L} = p_{1L} = p_{1L}^* x_1 = 0.15 \times 0.071 = 0.011$
  Water : $n_{2L} = p_{2L} = p_{2L}^* x_2 = 0.061 \times 0.93 = 0.057$

The condensed mole contents are
  Ethanol: $m_1 = n_{1H} - n_{1L} = 0.033 - 0.011 = 0.022$
  Water : $m_2 = n_{2H} - n_{2L} = 0.19 - 0.057 = 0.13$

Therefore, the mole fraction of ethanol in solution after condensation becomes
$$y_1 = \frac{m_1}{m_1 + m_2} = 0.14$$

The concentration in ABV unit is
$$ABV' = \frac{y_1}{0.304 + 0.696\ y_1} = 0.36$$

The condensed solution becomes ethanol solution of 36 v/v% concentration. This value is an ideal case, i.e., the initial condition of the surface inside box is originally dry (the left box in Fig. 4). There is water on the surface, i.e., the mucous membrane is 95 v/v% of water. The actual ethanol concentration will have time dependent Gaussian distribution along its depth. For more detail, we need to run numerical simulation.

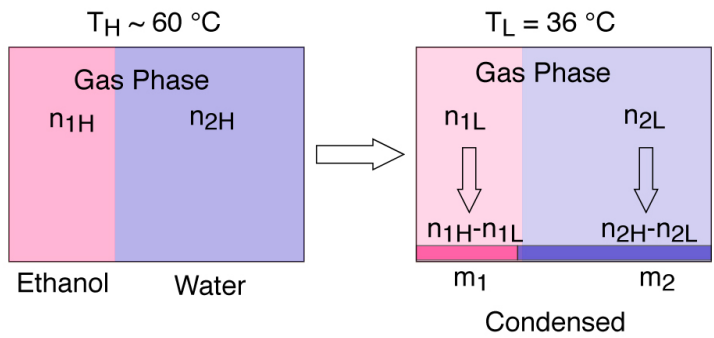

Fig. 4. Mass transfer in the condensation process.

5. Prompt Method
We use 40 v/v% whisky or similar alcohol, dripping on a gauze, inhale the vapor slowly at room temperature.

(1) Drip 1~2 ml (1~2 g weight ) of alcohol 40 v/v%, such as, whisky, on a gauze.
(2) Cover your nostrils with gauze (point where you drip the alcohol).(3) Slowly and carefully inhale the vapor. If one of your nostril has narrowing, block other one.
(4) After a few tens of seconds, the smell of the ethanol will disappear, then stop.

This method works well for the front part of the nasal cavity. This is suitable for clinical workers, because their working environment means that they need may prompt preventative measure at short notice.  Note that there is a risk of inflammation due to high concentration of alcohol, but we have to decide comparing with the risk from the corona virus itself. This method is only suggested for use by medical professionals, and is not considered appropriate for use by the general public.
 During the inhalation, the temperature of the alcohol drops to ~ 20°C due to heat of the vaporization. Therefore the vapor pressure becomes much lower than the heating method, and amount of the mass transfer (ethanol from gauze to nasal cavity) becomes roughly five times smaller. The ethanol concentration will be determined by the mass balance between adsorption and diffusion at the surface of mucous membrane. According to the rough estimate, it is around 30~40 v/v%. Further detail study is required.

6. Discussions
   The proposal discussed in this paper is still conceptual. We need further study on the following issues.
(1) At this moment on March 2020, the author does not have reliable data on the ethanol concentration required for disinfection of SARS-CoV-2. The author would like to ask collaborators to perform experiment on this virus as soon as possible in any laboratory capable of handling the virus safely.
(2) As discussed in Section. 2, the ethanol diffuses very quickly into water or coming out from solution, and evaporates quickly. There is also heat exchange due to evaporation. Therefore, we should be careful on how to evaluate actual ethanol concentration in the experiments.
(3) Generally speaking, the lipid becomes soft and active when it is warmed. Therefore, the above mentioned experiments should be performed at the body temperature of 36.5°C, not at a room temperature of 20°C.
(4) The side effects due to ethanol vapor inhalation are not yet evaluated, particularly, into the trachea and the lung. Careful clinical trials under medical doctor supervision will be necessary.
(5) Alcohol inhalation is equivalent to "smoking ethanol vapor". It will be required to discuss how to control applications to persons below the age of legal consumption of alcohol, and also for elderly people.
(6) Methanol($CH_3OH$) or methanol contained solution are highly prohibited to inhale, because of methanol toxicity.
(7) Once a day will be enough. It is better to perform this method at dinner time, and should not drive car right after this.

7. Conclusions
  The author discussed on possibility of disinfection of SARS-CoV-2 in human respiratory tract by ethanol vapor inhalation. The alcohol distillation helps to raise the ethanol concentration on the mucous membrane inside our recuperatory tract, where the viruses are suppose to remain and incubate. Because of nature of the lipid bilayer (i.e. it is lipophilic), there is a chance to disinfect the corona virus with alcohol vapor inhalation. The author would like to ask researchers in this field to do more works on this method.

Acknowledgment
The author would like to thank Dr. Masao Yamashita and Dr. Ryusuke Kuwahara for their various discussions and encouragements. The author also wishes to thank to Mr. Shuji Misumi and Mr. Andrean Hanley for their help on checking equations, and thank to Mr. Seita Taba for laboratory test of ethanol effect on E. coli. The author wishes to thank Dr. Cathal Cassidy for his careful reading of the manuscript, and thank to Ms. Yoko Shintani for her various help.

References
[1] Seymour S. Block, "Disinfection, Sterilization and Preservation", Fifth Edition, Lippincott Williams & Wilkins, 2001, ISBN o-683-30740-1, p239.
[2] Nobuji NODA et. al., "Virucidal Activity of Alcohols, Virucidal Efficiency of Alcohols Against Viruses in Liquid Phase", The Journal of the Japanese Association for Infection Diseases, Vol. 55, No.5, 1980.
[3] Akemi Shinmyo, "Effect of Density for Sterilization of Bacterias and Disinfection on Viruses" doctoral thesis (in Japanese), Tokyo Health Care University, 2019.
[4] WHO, Novel Coronavirus (2019-nCoV). Situation Report 50, WHO (2020)
[5] A to Z of Thermodynamics by Pierre Perrot. ISBN 0-19-856556-9.
[6] Rama Bansil and Brandley S. Turner, "The biology of mucus: Composition, synthesis and organization", Advanced Drug Delivery Reviews 124 (2018) 3-15.
[7] Ganong's Review of Medical Physiology (24 ed.). ISBN 0071780033.
[8] Zheng J, Wang YP, Dong Z, Yang ZQ, Sun W., "Nasal cavity volume and nasopharyngeal cavity volume in adults measured by acoustic rhinometry", US National Library of Medicine National Institutes of Health, 2000 No; 14 (11): 494-5
[9] Achim G. Beule, "Physiology and pathophysiology of respiratory mucosa of the nose and the paranasal sinuses", GMS Curr Top Otorhinolaryngol Head Neck Surg. 2010; 9: Doc07.
http://www.thcu.ac.jp/uploads/imgs/20190605090207.pdf

# Supplemental Information (Appendix)

A1. Alcohols Beverages and Dilution Ratio

The following alcohols can be used. Unfortunately, most of the red and white wines do not have high enough alcohol concentration. Liquors can be used by diluting with water. For example, Whisky single shot amount (~30 ml) adding water 30 ml will be easy.

|  | Alcohol Concentration ABV % | Dilution Alcohol : Water |
|---|---|---|
| Whiskey | 40% | 1:1 |
| Vodka | 40% | 1:1 |
| Gin | 40% | 1:1 |
| Chinese Baijiu | ~ 38% | 1:1 |
| 烧酒 | ~ 38% | 1:1 |
| Japanese Shochu | ~ 25% | 1:1/2 |
| Awamori Okinawa | ~30% | 1:1/2 |
| Fortified Wines (Sherry, Port, Madeira & Others) | 17 ~ 20% | as it is |
| Japanese Sake | 15~20% | higher than 16% is better |

Table A1. Alternative alcohols and optimum dilution with water.

A2. The Cup

The plastic and paper cup for coffee will be good for minimizing heat loss, thus good for our purpose. A red wine glass, with a broad bowl is the best.

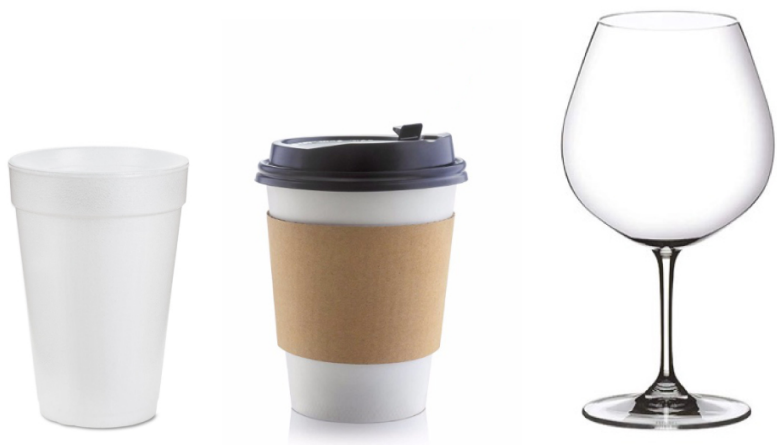

Plastic cups styrofoam made, paper cup with heat insulation, and a red wine glass.

A3. How to measure the volume of alcohols.
The measuring cup for kitchen is the best to measure the volume. The shot glass is also useful. Another good way is to measure the weight, where 100 ml = 100 g. The digital or mechanical weight scale for kitchen is useful. Place an empty cup on the weight scale, pour alcohol for 30 mg, then add 30 mg of hot water (1:1 dilution).

A4. How to heat.
(1) Whisky or others ~40 v/v%, refer the photos below.
   Boil water with a kettle. Pour the boiling water in an empty cup, or empty wine glass. Stop gas or electric power of a boiling kettle. Wait one minute, then the temperature of boiled water in the kettle become ~90°C (190°F). Empty the cup, pour a single shot or smaller amount of Whisky. Pour the same amount of the hot water of 90°C (190°F) from the kettle, you have 55~60°C (122~140°F) of alcohol solution, ~20% of ethanol concentration.

(2) Chinese Baijiu
   Same as Whisky.
   Or, dilute Baijiu with the same amount of water. Use a plastic and paper cup for coffee, heat in a microwave oven. You need roughly 500 Watt x 30 sec for 100 ml (~100 g), see below.

(3) Japanese sake or diluted Japanese Shochu.
   Use sake bottle, heat in hot water bath.
   Or heat with microwave oven. You need roughly 500 Watt x 30 sec for 100 ml (~100 g).

A5. Heating with microwave oven.
The microwave oven is a good way to heat. But, you need practice. The microwave power and required heating times are summarized in Table A4. You should try to heat with the tap water first, and measure the temperature with a thermometer. If temperature is higher than 60°C (140°F), reduce heating time.

| Total volume (weight) | Heating Power and Time from room temperature at 15°C (60°F) to 60 °C (140°F). | Heating Power and Time from temperature at 25°C (77°F) to 60 °C (140°F). |
| --- | --- | --- |
| 60 ml (60 g) | 500 W x 23 sec, 600 W x 19 sec | 500 W x 18 sec, 600 W x 15 sec |
| 100 ml (100 g) | 500 W x 38 sec, 600 W x 33 sec | 500 W x 30 sec, 600 W x 25 sec |

Required time can be estimated by $t = W \times dT \times 4.2/P$, where $t$ is time in seconds, $W$ is weight of the alcohol solution, $dT$ is heating temperature from the room temperature in °C, $P$ is power of the microwave oven in Watt. The cup itself need to be heated, so that you need prior test.

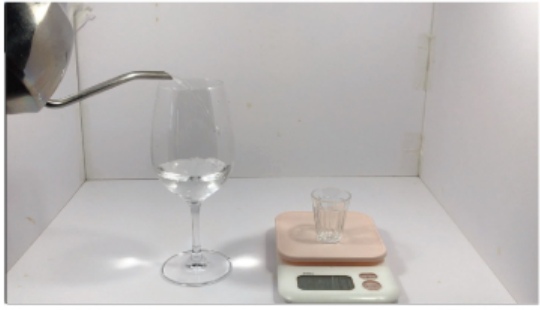
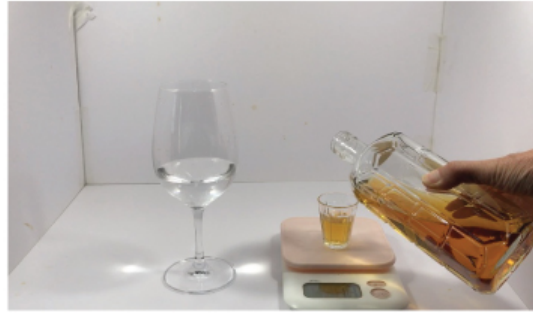
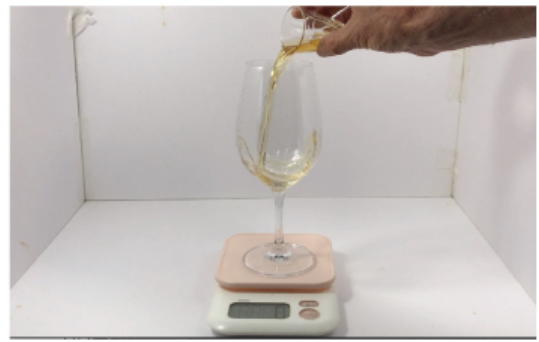
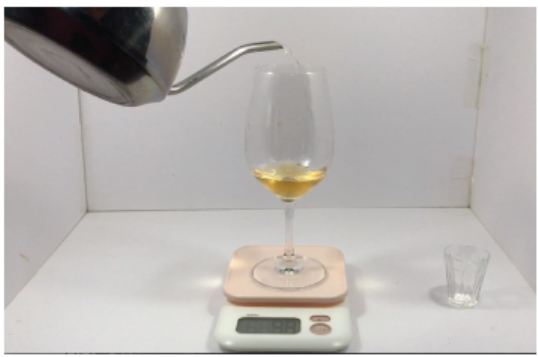

How to make the alcohol solution.

— *End of supplemental information* —